\documentclass[conference]{IEEEtran}
\usepackage[T1]{fontenc}
\usepackage[utf8]{inputenc}
\usepackage{amssymb}
\usepackage[pdftex]{graphicx}
\usepackage{subcaption} 
\usepackage{amsmath}
\usepackage{amsthm}
\usepackage{mathtools}
\usepackage{array}
\usepackage[noadjust]{cite}
\usepackage{algorithm}
\usepackage{algpseudocode}
\usepackage{xcolor}
\usepackage{tikz}
\usepackage{pgfplots}
\usepackage[hidelinks]{hyperref}
\usepackage[acronym]{glossaries}
\usepackage{dutchcal}
\pgfdeclarelayer{background}
\pgfdeclarelayer{foreground}
\pgfsetlayers{background,main,foreground}
\usepgfplotslibrary{fillbetween}
\usepgfplotslibrary{external}
\usetikzlibrary{spy,backgrounds}
\usepackage[binary-units=true]{siunitx}
\usepackage{balance}
\makeatletter
\let\MYcaption\@makecaption
\makeatother

\usepackage{subcaption}

\makeatletter
\let\@makecaption\MYcaption
\makeatother
\algnewcommand{\LineComment}[1]{\(\triangleright\) #1}

\DeclareSIUnit\frame{frame}
\DeclareSIUnit\PRB{PRB}

\newacronym{SNR}{SNR}{signal-to-noise ratio}
\newacronym{LDPC}{LDPC}{low-density parity-check}
\newacronym{QAM}{QAM}{quadrature amplitude modulation}
\newacronym{QPSK}{QPSK}{quadrature phase-shift keying}
\newacronym{BICM}{BICM}{bit-interleaved coded modulation}
\newacronym{BCE}{BCE}{binary cross-entropy}
\newacronym{LLR}{LLR}{log-likelihood ratio}
\newacronym{BMI}{BMI}{bit-wise mutual information}
\newacronym{KL}{KL}{Kullback–Leibler}
\newacronym{BMD}{BMD}{bit-metric decoding}
\newacronym{BER}{BER}{bit error rate}
\newacronym{NN}{NN}{neural network}
\newacronym{GS}{GS}{geometric shaping}
\newacronym{SIP}{SIP}{superimposed pilot}
\newacronym{iid}{i.i.d.\@}{independent and identically distributed}
\newacronym{SGD}{SGD}{stochastic gradient descent}
\newacronym{wrt}{w.r.t.\@}{with respect to}
\newacronym{MAP}{MAP}{maximum a posteriori}
\newacronym{LMMSE}{LMMSE}{linear minimum mean square error}
\newacronym{AWGN}{AWGN}{additive white Gaussian noise}
\newacronym{RBF}{RBF}{Rayleigh block fading}
\newacronym{OFDM}{OFDM}{orthogonal frequency division multiplexing}
\newacronym{3GPP}{3GPP}{3rd Generation Partnership Project}
\newacronym{5GNR}{5G NR}{5G New Radio}
\newacronym{PRB}{PRB}{physical resource block}
\newacronym{CSI}{CSI}{channel state information}
\newacronym{MSE}{MSE}{mean squared error}
\newacronym{DC}{DC}{direct current}
\newacronym{PAPR}{PAPR}{peak-to-average power ratio}
\newacronym{DMRS}{DMRS}{demodulation reference signal}
\newacronym{TTI}{TTI}{transmission time interval}
\newacronym{CDF}{CDF}{cumulative distribution function}
\newacronym{RE}{RE}{resource element}
\newacronym{MIMO}{MIMO}{multiple-input multiple-output}
\newacronym{IDFT}{IDFT}{inverse discrete Fourier transform}
\newacronym{DFT}{DFT}{discrete Fourier transform}
\newacronym{CP}{CP}{cyclic prefix}
\newacronym{FFT}{FFT}{fast Fourier transform}
\newacronym{ISI}{ISI}{inter-symbol interference}
\newacronym{ICI}{ICI}{inter-carrier interference}
\newacronym{LOS}{LoS}{line-of-sight}
\newacronym{NLOS}{NLoS}{non-line-of-sight}
\renewcommand{\vec}[1]{\mathbf{#1}}

\newcommand{\gv}{\vec{g}}

\newcommand{\pv}{\vec{p}}

\newcommand{\sv}{\vec{s}}

\newcommand{\wv}{\vec{w}}
\newcommand{\xv}{\vec{x}}
\newcommand{\yv}{\vec{y}}
\newcommand{\zv}{\vec{z}}
\newcommand{\zerov}{\vec{0}}

\newcommand{\Fm}{\vec{F}}
\newcommand{\Gm}{\vec{G}}
\newcommand{\Hm}{\vec{H}}
\newcommand{\Id}{\vec{I}}

\newcommand{\Pm}{\vec{P}}

\newcommand{\Rm}{\vec{R}}
\newcommand{\Sm}{\vec{S}}

\newcommand{\Xm}{\vec{X}}
\newcommand{\Ym}{\vec{Y}}
\newcommand{\Zm}{\vec{Z}}

\newcommand{\Cc}{{\cal C}}

\newcommand{\Lc}{{\cal L}}

\newcommand{\Nc}{{\cal N}}

\newcommand{\CC}{\mathbb{C}}

\newcommand{\RR}{\mathbb{R}}

\newcommand{\htp}{^{\mathsf{H}}}

\newcommand{\LB}{\left(}
\newcommand{\RB}{\right)}
\newcommand{\LP}{\left\{}
\newcommand{\RP}{\right\}}
\newcommand{\LSB}{\left[}
\newcommand{\RSB}{\right]}

\renewcommand{\ln}[1]{\mathop{\mathrm{ln}}\LB #1\RB}
\renewcommand{\log}[1]{\mathop{\mathrm{log}}\LB #1\RB}
\renewcommand{\exp}[1]{\mathop{\mathrm{exp}}\LB #1\RB}

\newcommand{\EE}{{\mathbb{E}}}

\newcommand\abs[1]{\left| #1 \right|}

\begin{document}
\title{Trimming the Fat from OFDM: Pilot- and CP-less Communication with End-to-end Learning}

\IEEEoverridecommandlockouts 

\author{\IEEEauthorblockN{Fay\c{c}al Ait Aoudia and Jakob Hoydis}
\IEEEauthorblockA{Nokia Bell Labs, Paris, France\\
 \{faycal.ait\_aoudia, jakob.hoydis\}@nokia-bell-labs.com}
 \thanks{Part of this work was submitted to a journal for publication (preprint arXiv:2009.05261).}
 }

\maketitle

\begin{abstract} \Gls{OFDM} is one of the dominant waveforms in wireless communication systems due to its efficient implementation.
However, it suffers from a loss of spectral efficiency as it requires a \gls{CP} to mitigate \gls{ISI} and pilots to estimate the channel.
We propose in this work to address these drawbacks by learning a \gls{NN}-based receiver jointly with a constellation geometry and bit labeling at the transmitter, that allows \gls{CP}-less and pilotless communication on top of \gls{OFDM} without a significant loss in \gls{BER}. Our approach enables at least \SI{18}{\percent} throughput gains compared to a pilot and \gls{CP}-based baseline, and at least \SI{4}{\percent} gains compared to a system that uses a neural receiver with pilots but no \gls{CP}.

\end{abstract}
\glsresetall

\section{Introduction}

The next generation of cellular communication systems is expected to meet throughput, delay, and latency requirements beyond what is supported by current radio infrastructures~\cite{9040431}.
Most of today's wireless systems rely on \gls{OFDM}, which is used in 4G, WiFi, and 5G.
The success of \gls{OFDM} is due to its very efficient implementation, owing to the \gls{FFT} algorithm, easy single-tap equalization at the receiver, as well as granular access to the time and frequency resource grid.
However, the \gls{OFDM} waveform is not flawless. It suffers from high \gls{PAPR}, poor spectral containment, high sensitivity to Doppler spread, and loss of spectral efficiency due to the use of a \gls{CP} to mitigate \gls{ISI} and pilot signals to estimate the channel.

This work addresses this latter weakness of \gls{OFDM} by leveraging end-to-end learning to design a \gls{CP}-less and pilotless \gls{OFDM} communication system.
The key idea of end-to-end learning is to implement the transmitter, channel, and receiver as a single \gls{NN}, referred to as an autoencoder, that is trained to achieve the highest possible information rate~\cite{9118963}.
Since its first application to wireless communications~\cite{8054694}, end-to-end learning has been extended to other fields including optical wireless~\cite{8819929} and optical fiber~\cite{Karanov:18}.
We propose in this work to  learn a constellation and associated bit labeling which are used to modulate coded bits on all \glspl{RE}.
The learned constellation is forced to have zero mean to avoid an unwanted \gls{DC} offset.
As a side-effect, it can not be interpreted as a constellation with superimposed pilots.
Moreover, to maximize spectral efficiency, no \gls{CP} is used in the system, similarly to~\cite{8052521}, in which pilots were however still leveraged.
On the receiver side, a \gls{NN} is jointly optimized with the constellation and bit labeling to compute \glspl{LLR} for the transmitted bits directly from the post-\gls{FFT} received signal.

For benchmarking, we have implemented a strong baseline receiver which relies on \gls{LMMSE} channel estimation with perfect tempo-spectral covariance matrix knowledge, and pilot patterns from \gls{5GNR}.
Our results demonstrate that the proposed approach achieves \glspl{BER} similar or lower than the ones achieved by the baseline, and similar to the ones achieved by an \gls{NN}-based receiver with \gls{QAM} and pilots.
As a consequence, end-to-end learning enables additional throughput gains of at least \SI{18}{\percent} compared to the \gls{LMMSE}-based baseline, and of at least \SI{4}{\percent} compared to a system leveraging a neural receiver and \gls{QAM}, pilots, but no \gls{CP}.

\paragraph*{Notations}
Boldface upper-case (lower-case) letters denote matrices (column vectors);
regular lower-case letters denote scalars.
$\RR$ ($\CC$) is the set of real (complex) numbers; 
$()^*$ is the complex conjugate operator.
$\ln{\cdot}$ denotes the natural logarithm and $\log{\cdot}$ the binary logarithm.
The $(i,k)$ element of a matrix $\Xm$ is denoted by $X_{i,k}$. The $k^{th}$ element of a vector $\xv$ is $x_k$  and $\text{diag}(\xv)$ is a diagonal matrix with $\xv$ as diagonal.
The operators $()\htp$ and $\text{vec}()$ denote the Hermitian transpose and vectorization, respectively. 
For two matrices $\Xm$ and $\Ym$, the Kronecker product is denoted by $\Xm \otimes \Ym$.
Finally, $\Id_n$ denotes the identity matrix of size $n$, $\Fm_n$ denotes the unitary \gls{DFT} matrix of size $n$, and $\zerov_{n\times m}$ the zero matrix of size $n \times m$.

\section{Channel model and baseline}
\label{sec:sys}
This section introduces the channel model and details a baseline receiver algorithm for performance benchmarking with the machine learning-based approaches.

\begin{figure}
	\centering
	\includegraphics[width=0.48\textwidth]{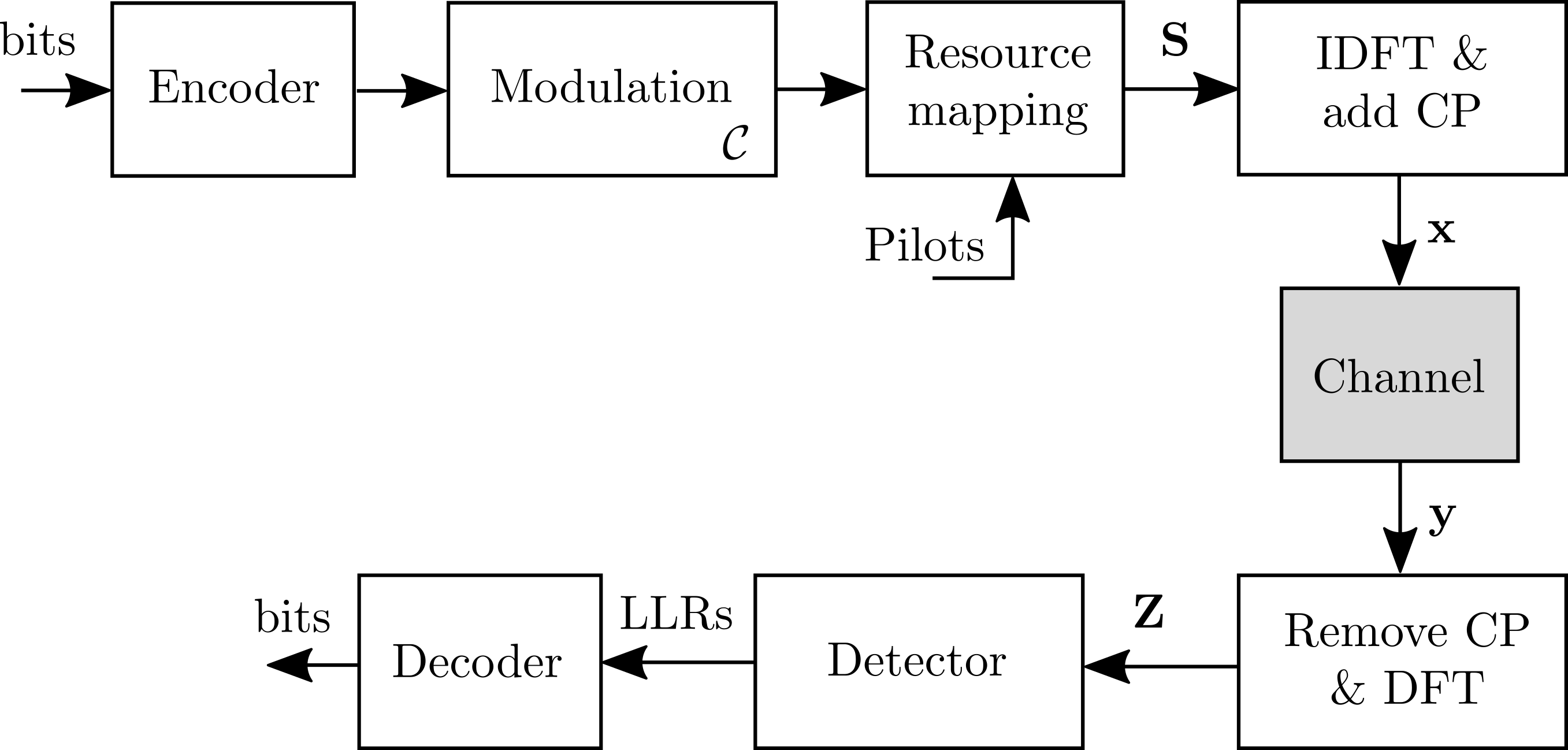}
	\caption{OFDM-based communication system. Interleaving and deinterleaving are not shown for compactness.\label{fig:ofdm_sys}}
\end{figure}

\subsection{Channel model}
\label{sec:ch_mod}

An \gls{OFDM} system is considered and the time-frequency matrix of symbols to transmit is denoted by $\Sm \in \CC^{n_S \times n_T}$, where $n_S$ is the number of subcarriers and $n_T$ the number of OFDM symbols forming the resource grid.
Depending on the transmission scheme, elements of $\Sm$ can be either a pilot symbol or a modulated symbol according to a constellation $\Cc$, e.g., \gls{QAM} or a learned constellation.
As illustrated in Fig.~\ref{fig:ofdm_sys}, the transmitted signal $\xv$ is obtained by applying the \gls{IDFT} to $\Sm$ and adding a \gls{CP}, which can be formally written as
\begin{equation}
	\xv = \text{vec}\LB \mathbf{C_T} \Fm_{n_S}\htp \Sm \RB
\end{equation}
where $\mathbf{C_T} \in \CC^{(n_S  + n_{CP}) \times n_S}$ is the operator corresponding to the addition of the \gls{CP}, $n_{CP}$ being the \gls{CP} length.
More precisely,
\begin{equation}
	\mathbf{C_T} \coloneqq
	\left[
	\begin{array}{c}
		\begin{array}{cc}
			\zerov_{n_{CP} \times (n_S - n_{CP})}	& \Id_{n_{CP}}
		\end{array} \\
		\Id_{n_S}
	\end{array}
	\right].
\end{equation}
The scheme presented in Section~\ref{sec:e2e} does not require the use of a \gls{CP}, i.e., $n_{CP} = 0$ and $\mathbf{C_T} = \Id_{n_S}$.

We consider a multi-tap and time varying channel with $n_R$ taps.
Elements of the channel output $\yv \in \CC^{(n_S+n_{CP})n_T}$ are given by
\begin{equation}
	\label{eq:taps_channel}
	y_t = \sum_{i=0}^{n_R- 1} x_{t-i} h_{i,t} + w_t
\end{equation}
where $h_{i,t} \in \CC$ is the channel coefficient for the $i^{\text{th}}$ tap at time step $t$, and $w_t \in \CC$ the additive white complex Gaussian noise with variance $\sigma^2$.
The transmitted symbols are assumed to have an average energy equal to one, i.e., $\EE\LP \abs{x_t}^2 \RP = 1$.

At the receiver, the signal $\yv$ is first reshaped as a matrix $\Ym \in \CC^{(n_S + n_{CP}) \times n_T}$.
The \gls{CP} is then removed and the \gls{DFT} is applied to obtain
\begin{equation}
	\label{eq:Z}
	\Zm = \Fm_{n_S} \mathbf{C_R} \text{vec}^{-1} \LB \yv \RB
\end{equation}
where $\mathbf{C_R} \coloneqq \LSB \zerov_{n_S \times n_{CP}} \Id_{n_S} \RSB$.
If no \gls{CP} is used, then $\mathbf{C_R} = \Id_{n_S}$.
The effective transfer function in the time-frequency domain is
\begin{equation}
	\label{eq:ofdm_ch}
	\zv = \Gm \sv + \wv
\end{equation}
where $\zv \coloneqq \text{vec}\LB \Zm \RB$, $\sv \coloneqq \text{vec}\LB \Sm \RB$, and $\wv$ is a vector of white Gaussian noise with variance $\sigma^2$ per element.
The channel matrix $\Gm \in \CC^{n \times n}$, with $n \coloneqq n_S n_T$, is
\begin{equation}
	\Gm = \LB \Id_{n_T} \otimes \LB \Fm_{n_S}\mathbf{C_R} \RB \RB \Hm \LB \Id_{n_T} \otimes \LB \mathbf{C_T}\Fm_{n_S}\htp \RB \RB
\end{equation}
where $\Hm \in \CC^{n_T \LB n_{CP} + n_S \RB \times n_T \LB n_{CP} + n_S \RB}$, is the channel matrix in time domain composed of the channel taps.

Because a time-varying channel is assumed, $\Gm$ is non-diagonal in general.
Therefore, we rewrite the transfer function~\eqref{eq:ofdm_ch} as
\begin{equation}
	\label{eq:ofdm_ch_2}
	\zv = \text{diag}\LB \gv \RB \sv + \LB \Gm - \text{diag}\LB \gv \RB \RB \sv + \wv
\end{equation}
where $\gv \in \CC^n$ is the diagonal of $\Gm$.
The first term on the right-hand side of~\eqref{eq:ofdm_ch_2} corresponds to the single-tap coefficients, whereas the second term accounts for the \gls{ICI} due to the time-varying channel coefficients, as well as for \gls{ISI} if $n_{CP} \leq n_R - 1$.

\begin{figure}
	\centering
	\includegraphics[width=0.28\textwidth]{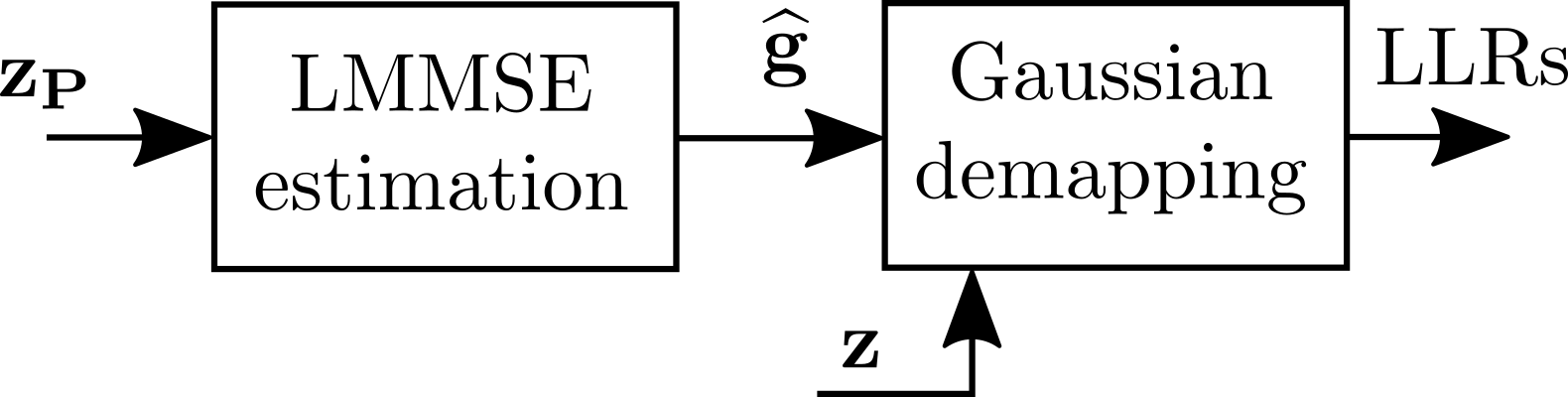}
	\caption{Architecture of the detector baseline.\label{fig:baseline}}
\end{figure}

As shown in Fig.~\ref{fig:ofdm_sys}, $\Zm$ is processed by a detector that computes \glspl{LLR} for the transmitted coded bits, which are then fed to a channel decoder that reconstructs the transmitted bits.
The rest of this section presents a baseline for such a detector, which performs \gls{LMMSE} channel estimation based on transmitted pilots, followed by soft-demapping assuming Gaussian noise, as illustrated in Fig.~\ref{fig:baseline}.

\subsection{LMMSE channel estimation and Gaussian demapping}
\label{sec:bsl}

The considered baseline assumes a single-tap time-frequency channel, i.e., $\LB \Gm - \text{diag}\LB \gv \RB \RB = \zerov_{n \times n}$, so that
\begin{equation}
	\label{eq:single_tap_transf}
	\zv = \text{diag}\LB \gv \RB \sv + \wv.
\end{equation}
This is equivalent to assuming no \gls{ICI}.
Note that this is neither true in practical channels nor in the channel model depicted in Section~\ref{sec:ch_mod}.
However, this is a quite usual assumption as it enables convenient single-tap equalization.
Let us denote by $\Rm \in \CC^{n \times n}$ the covariance matrix of $\gv$.
The matrix $\Rm$ determines the temporal and spectral correlation of the OFDM channel.

We denote by $\Pm\in\CC^{n_S \times n_T}$ the pilot matrix, whose entry $P_{i,k}$ is zero if the \gls{RE} on the $i^{th}$ subcarrier and $k^{th}$ time slot is carrying data, or equal to the pilot value otherwise.
Let $n_P$ be the number of pilot-carrying \glspl{RE} which are considered for channel estimation.
For these \glspl{RE}, we can re-write~\eqref{eq:single_tap_transf} considering only the pilots in vectorized form as
\begin{equation}
\mathbf{z_P} = \mathbf{\Pi} \LB \text{diag}(\pv) \gv + \wv\RB
\end{equation}
where $\pv \coloneqq \text{vec}(\Pm)$, and $\mathbf{\Pi}$ is the $n_P \times n$ matrix which selects only the elements carrying pilot symbols.
Assuming knowledge of $\Rm$ at the receiver, the \gls{LMMSE} channel estimate is (e.g.,~\cite[Lemma B.17]{massivemimobook}) 
\begin{multline}
\label{eq:lmmse_est}
\widehat{\gv} = \Rm \text{diag}(\pv)\htp \mathbf{\Pi}\htp \cdot\\
	\LB \mathbf{\Pi} \LB \text{diag}(\pv) \Rm \text{diag}(\pv)\htp + \sigma^2 \Id_n \RB \mathbf{\Pi}\htp \RB^{-1} \mathbf{z_p}.
\end{multline}
Using this result, the received signal, which incorporates both pilots and data, can be re-written in vector form as
\begin{equation}
\zv = \text{diag}(\widehat{\gv})\sv + \underbrace{\text{diag}(\widetilde{\gv})\sv + \wv}_{ \coloneqq \widetilde{\wv}}
\end{equation}
where $\widetilde{\gv} = \gv - \widehat{\gv}$ is the channel estimation error with correlation matrix $\widetilde{\Rm}\in\CC^{n\times n}$, given as
\begin{multline}
\label{eq:est_err}
\widetilde{\Rm} = \EE \LP \widetilde{\gv}\widetilde{\gv}\htp \RP = \Rm - \Rm \text{diag}(\pv)\htp \mathbf{\Pi}\htp\\
\LB \mathbf{\Pi} \LB \text{diag}(\pv) \Rm \text{diag}(\pv)\htp + \sigma^2 \Id_n \RB \mathbf{\Pi}\htp \RB^{-1} \mathbf{\Pi} \text{diag}(\pv)\Rm
\end{multline}
and $\widetilde{\wv}$ is the sum of noise and residual interference due to imperfect channel estimation.

Soft-demapping is performed assuming that $\widetilde{\wv}$ is Gaussian.\footnote{This is typically not true as $\text{diag}(\widetilde{\gv})\sv$ is not Gaussian.}
Let us denote by $m$ the number of bits per channel use, by $\Cc = \{c_1,\dots,c_{2^m}\}$ the constellation, and by $\Cc_{i,0}$ ($\Cc_{i,1}$) the subset of $\Cc$ which contains all constellation points with the $i^{th}$ bit label set to 0 (1).
The \gls{LLR} for the $i^{th}$ bit ($i \in \{1,\dots,m\}$) of the $k^{th}$ \gls{RE} ($k \in \{1,\dots,n\}$) is computed as follows:
\begin{equation}
\text{LLR}(k,i) = \ln{\frac{\sum_{c \in \Cc_{i,1}} \exp{-\frac{1}{\widetilde{\sigma}_k^2} \abs{z_k - \widehat{g}_k c}^2}}{\sum_{c \in \Cc_{i,0}} \exp{-\frac{1}{\widetilde{\sigma}_k^2} \abs{z_k - \widehat{g}_kc}^2}}}
\end{equation}
where $\widetilde{\sigma}_k^2 = \EE \LP \widetilde{w}_k\widetilde{w}_k^* \RP = \widetilde{R}_{k,k} + \sigma^2$  for $k=1,\dots, n$. Whenever $k$ corresponds to the index of a pilot symbol, no \gls{LLR} value is computed. After deinterleaving, the \glspl{LLR} are fed to a channel decoding algorithm (e.g., belief propagation) which computes predictions of the transmitted bits.

\section{End-to-end learning for OFDM}
\label{sec:e2e}

End-to-end learning of communication systems~\cite{8054694} consists in implementing a transmitter, channel, and receiver as a single \gls{NN} referred to as an autoencoder, and jointly optimizing the trainable parameters of the transmitter and receiver for a specific channel model.
In our setting, training aims at minimizing the total \gls{BCE} [\si{\bit\per\frame}], defined as
\begin{equation}
\label{eq:bce}
\Lc \coloneqq -\sum_{k \in \Nc_D}\sum_{i=1}^m \EE_{b_{k,i}, \zv} \LP \log{Q_{k,i}\LB b_{k,i}|\zv \RB }\RP
\end{equation}
where $\Nc_D$ is the set of size $n_D$ of indexes of \glspl{RE} carrying data symbols, $b_{k,i}$ is the $i^{th}$ bit transmitted in the $k^{th}$ resource element, and $Q_{k,i}\LB \cdot | \zv \RB$ is the receiver estimate of the posterior distribution of the $i^{th}$ bit transmitted in the $k^{th}$ \gls{RE}, given the received signal after \gls{OFDM} demodulation $\zv$.
Since \eqref{eq:bce} is numerically difficult to compute, it is estimated through Monte Carlo sampling as
\begin{equation}
\label{eq:bce_mc}
\Lc \approx -\frac{1}{S} \sum_{l=1}^S \sum_{k \in \Nc_D}\sum_{i=1}^m \LP \log{Q_{k,i}\LB b_{k,i}^{[l]}|\zv^{[l]} \RB }\RP
\end{equation}
where $S$ is the batch size, i.e., the number of samples used to estimate $\Lc$, and the superscript $[l]$ is used to refer to the $l^{th}$ sample within a batch.
Interestingly, as proved in~\cite{arXiv.2009.05261}, minimzing $\Lc$ is equivalent to maximizing an achievable rate, assuming that a mismatched \gls{BMD} receiver is used.

Training of the transmitter typically involves joint optimization of the constellation geometry and bit labeling~\cite{9118963}.
Such an approach is adopted in this work and extended to \gls{OFDM} channels in Section~\ref{sec:e2e_gs}.
Moreover, a centered constellation is learned, assuming that no pilots are transmitted and no \gls{CP} is used, as shown in Fig.~\ref{fig:sys_nn}.
This avoids an unwanted \gls{DC} offset and removes the throughput loss due to the transmission of reference signals and \gls{CP} that carry no data.
The architecture of the \gls{NN}-based receiver is detailed in Section~\ref{sec:nn_arch}.

\subsection{Learning of geometric shaping and bit-labeling}
\label{sec:e2e_gs}

\begin{figure}
	\centering
	\includegraphics[width=0.48\textwidth]{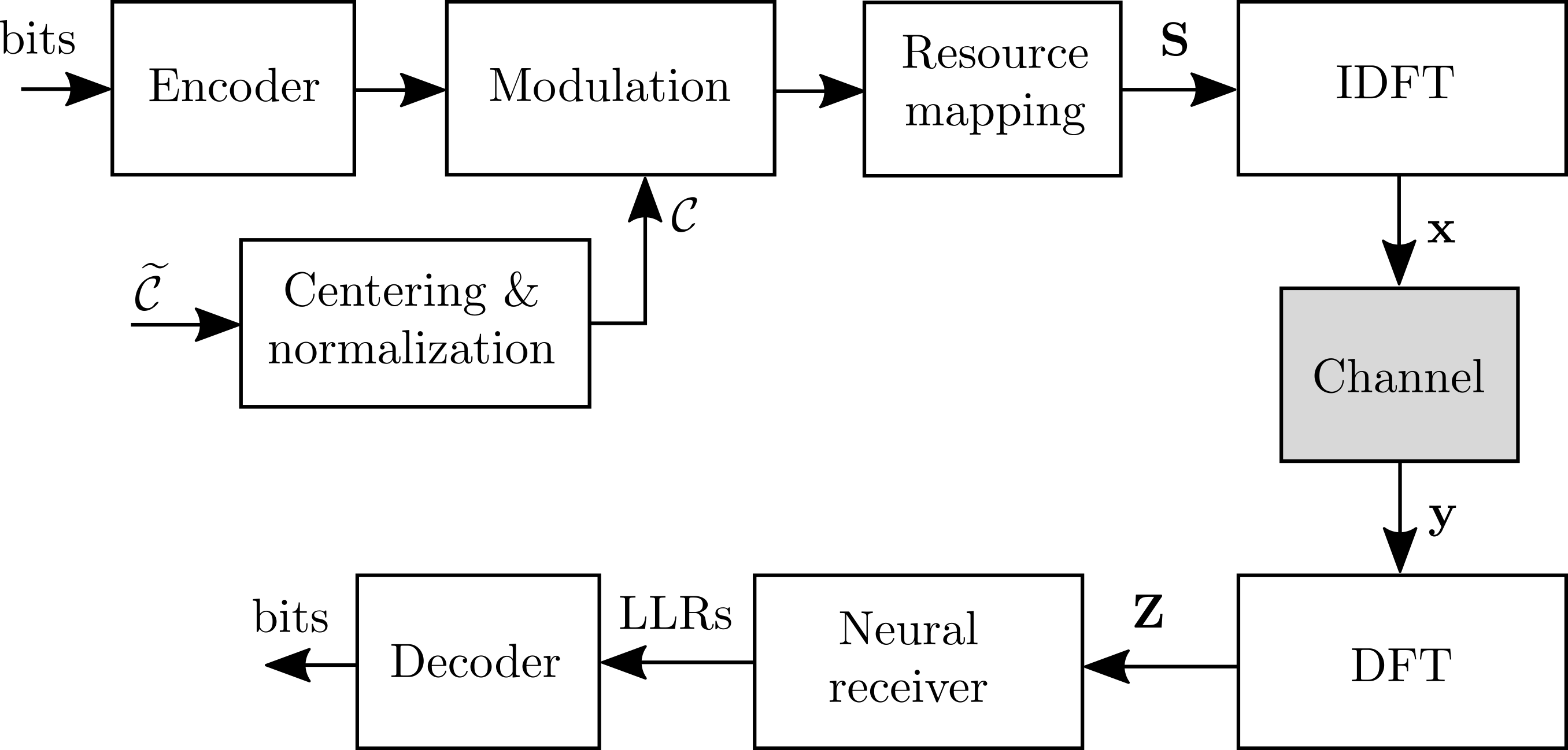}
	\caption{End-to-end learning system.\label{fig:sys_nn}}
\end{figure}

The system architecture we have adopted in this work is depicted in Fig.~\ref{fig:sys_nn}.
On the transmitter side, the trainable parameters consist of a set denoted by $\widetilde{\Cc}$ of $2^m$ complex numbers corresponding to the constellation points.
The constellation used for transmitting data is obtained by centering and normalizing $\widetilde{\Cc}$, i.e.,
\begin{equation}
\label{eq:const}
\Cc = \frac{\widetilde{\Cc} - \frac{1}{2^m}\sum_{c \in \widetilde{\Cc}}c}{\sqrt{\frac{1}{2^m}\sum_{c \in \widetilde{\Cc}}\abs{c}^2 - \abs{\frac{1}{2^m}\sum_{c \in \widetilde{\Cc}}c}^2}}.
\end{equation}
Normalization of the constellation ensures that it has unit average power, while  centering forces the constellation to have zero mean and prevents learning of embedded superimposed pilots.
As no pilots are transmitted, the receiver can only exploit the constellation geometry to reconstruct the transmitted bits and mitigate \glspl{ISI} due to the lack of \gls{CP} and \glspl{ICI} due to Doppler spread.
Compared to previous work such as~\cite{9118963}, a single constellation is learned, which is used for all \glspl{SNR}, Doppler, and delay spreads.
On the receiver side, an \gls{NN} that operates on multiple subcarriers and \gls{OFDM} symbols is leveraged, whose architecture is detailed in Section~\ref{sec:nn_arch}.
As in~\cite{9118963}, training of the end-to-end systems is done on the total \gls{BCE} estimated by~(\ref{eq:bce_mc}).

\subsection{Receiver architecture}
\label{sec:nn_arch}

\begin{figure}
     \begin{subfigure}[b]{\columnwidth}
         \centering
		\includegraphics[width=0.8\textwidth]{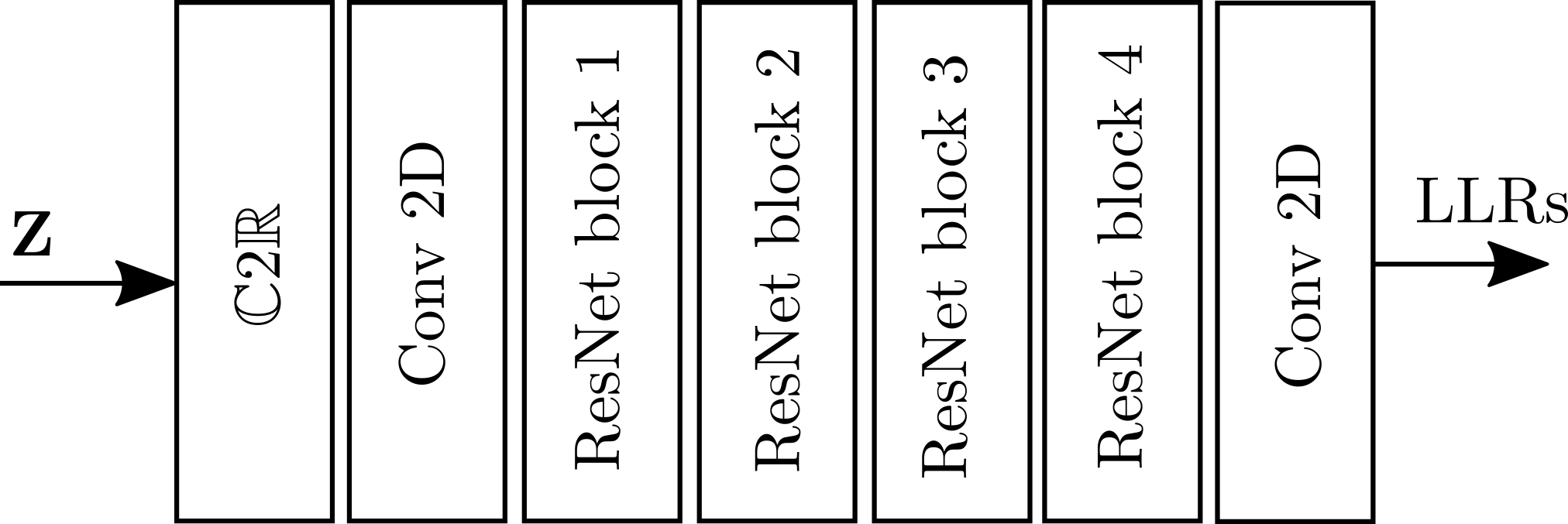}
         \caption{Architecture of the neural receiver.}
     \end{subfigure}

     \begin{subfigure}[b]{\columnwidth}
         \centering
         \includegraphics[width=0.8\textwidth]{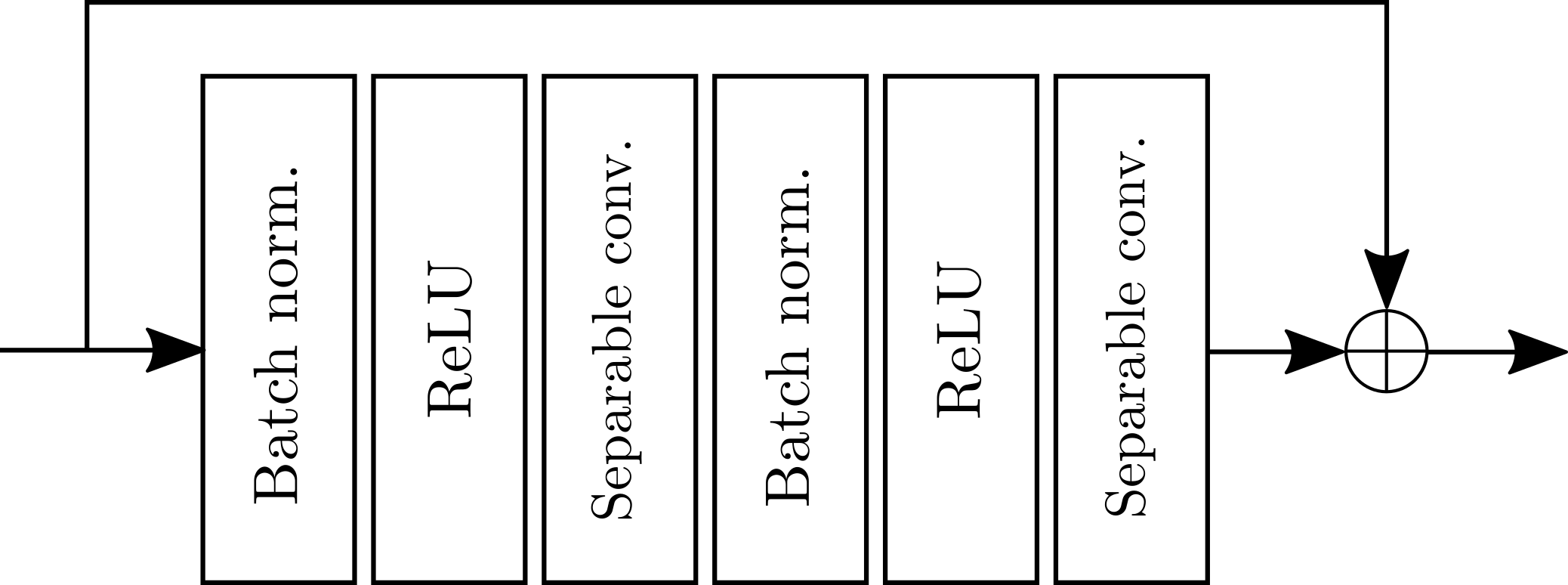}
         \caption{Architecture of a ResNet block.\label{fig:resnet}}
     \end{subfigure}
        \caption{The receiver is implemented by a residual convolutional neural network.}
        \label{fig:nn_rx}	
\end{figure}

The architecture of the \gls{NN} implementing the receiver is shown in Fig.~\ref{fig:nn_rx}.
It is a convolutional residual \gls{NN}~\cite{He_2016_CVPR} that takes as input the received baseband channel samples after the \gls{DFT} $\Zm$ of dimension $n_S \times n_T$, and outputs a 3-dimensional tensor of \glspl{LLR} of dimension $n_S \times n_T \times m$ that is fed to the channel decoder.
The \gls{NN} hence substitutes the estimator and demapper shown in Fig.~\ref{fig:baseline}.
The first layer $\CC2\RR$ converts the complex-valued input tensor of dimension $n_S \times n_T$ into a real-valued 3-dimensional tensor of dimension $n_S \times n_T \times 2$ by stacking the real and imaginary parts into an additional dimension.
Separable convolutional layers are used to reduce the number of weights, without incurring significant loss of performance.
Table~\ref{tab:nn_rx} provides details on the \gls{NN} implementing the receiver.
All convolutional layers use zero-padding to ensure that the dimensions of the output are the same as the ones of the input.
Dilation is leveraged to increase the receptive field of the convolutional layers.
A similar but somewhat larger architecture was used in~\cite{deeprx}.

\begin{table}
\begin{center}
  \begin{tabular}{ l | c | c | c }
    \hline
    Layer			& Channels	& Kernel size 	& Dilatation rate	\\ \hline
    Input Conv2D 	& 256		& (3,3)			& (1,1)				\\ \hline
    ResNet block 1	& 256		& (3,3)			& (3,1)				\\ \hline
    ResNet block 2	& 256		& (3,3)			& (6,2)				\\ \hline
    ResNet block 3	& 256		& (3,3)			& (6,2)				\\ \hline
    ResNet block 4	& 256		& (3,3)			& (3,1)				\\ \hline
    Output Conv2D 	& $m$		& (1,1)			& (1,1)				\\ \hline
  \end{tabular}
\end{center}
\caption{Architecture details of the \gls{NN} implementing the receiver.\label{tab:nn_rx}}
\end{table}

\begin{figure}
	\centering
	
     \begin{subfigure}[b]{0.45\columnwidth}
         \centering
		\includegraphics[width=0.9\textwidth]{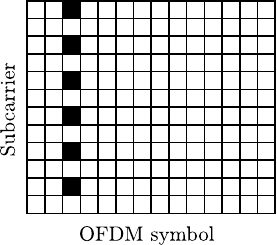}
         \caption{Pilot pattern ``1P''.\label{fig:pp1}}
     \end{subfigure}
     \begin{subfigure}[b]{0.45\columnwidth}
         \centering
		\includegraphics[width=0.9\textwidth]{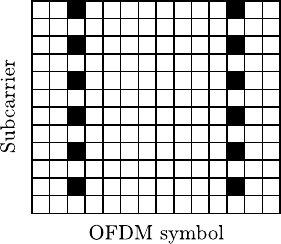}
         \caption{Pilot pattern ``2P''.\label{fig:pp2}}
     \end{subfigure}
     \caption{Pilot patterns from \gls{5GNR} used in simulations.\label{fig:pp}}
\end{figure}

\section{Simulation results}
\label{sec:results}

We will now present the results of the simulations we have conducted to evaluate the machine learning-based scheme introduced in the previous section, referred to as the \gls{GS} scheme.
We start by explaining the training and evaluation setup.
The \gls{GS} scheme is then compared to several other baselines.

\begin{figure}
	\centering
	\includegraphics[width=0.4\textwidth]{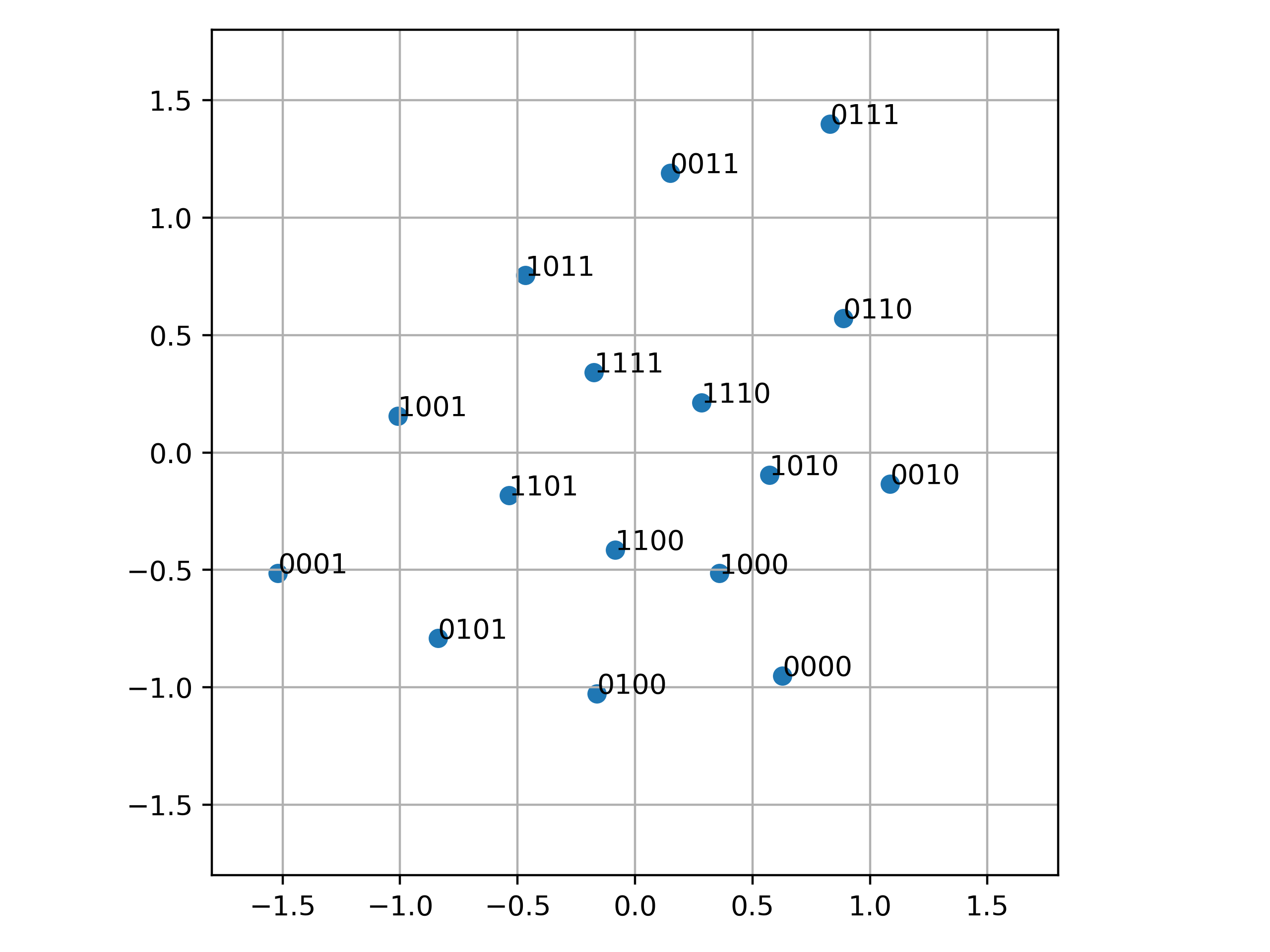}
	\caption{Learned constellation geometry and labeling.\label{fig:const}}
\end{figure}

\subsection{Evaluation setup}
\label{sec:setup}

\begin{table}
\begin{center}
  \begin{tabular}{ l | c | c }
    \hline
    Parameter			& Symbol (if any) & Value			\\ \hline
    Number of OFDM symbols		&	$n_T$ 		& 14 (1 slot)	\\ \hline
    Number of subcarriers 		&	$n_S$		& 72 (6 PRBs)	\\ \hline
    Frequency carrier 			&(None)		& \SI{3.5}{\GHz}	\\ \hline
    Subcarrier spacing			&	(None)	& \SI{30}{\kHz}	\\ \hline
    Cycle prefix duration		&	$n_{CP}$	& 	\multicolumn{1}{p{2cm}}{6 symbols\newline(\gls{5GNR} \gls{CP} of \SI{2.34}{\micro\second})}	\\ \hline
    Channel models				&	(None)	&	\multicolumn{1}{p{2cm}}{3GPP-3D\newline LoS and NLoS}\\ \hline
    Number of taps				&	$n_R$	&	5 \\ \hline
    Learning rate				&	(None)		& $10^{-3}$		\\ \hline
    Batch size for training		&	$S$			& 100 frames	\\ \hline
    Bit per channel use			&	$m$		&	\SI{4}{\bit}	\\ \hline
    Code length					&	(None)		& \SI{1024}{\bit} \\ \hline
    Code rate					&	$r$			& $\frac{2}{3}$ \\ \hline
    Speed range used for training &		(None)		& \multicolumn{1}{p{2cm}}{$0$ to\newline\SI{130}{\km\per\hour}}	\\ \hline
  \end{tabular}
\end{center}
\caption{Parameters used for training and evaluation\label{tab:setup_param}}
\end{table}

\begin{figure*}
	\centering
	
     \begin{subfigure}[b]{\linewidth}
         \centering
		\includegraphics[width=1.0\textwidth]{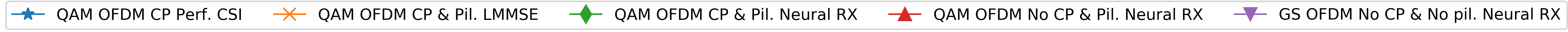}
     \end{subfigure}
	
     \begin{subfigure}[b]{0.3\linewidth}
         \centering
		\includegraphics[width=1.0\textwidth]{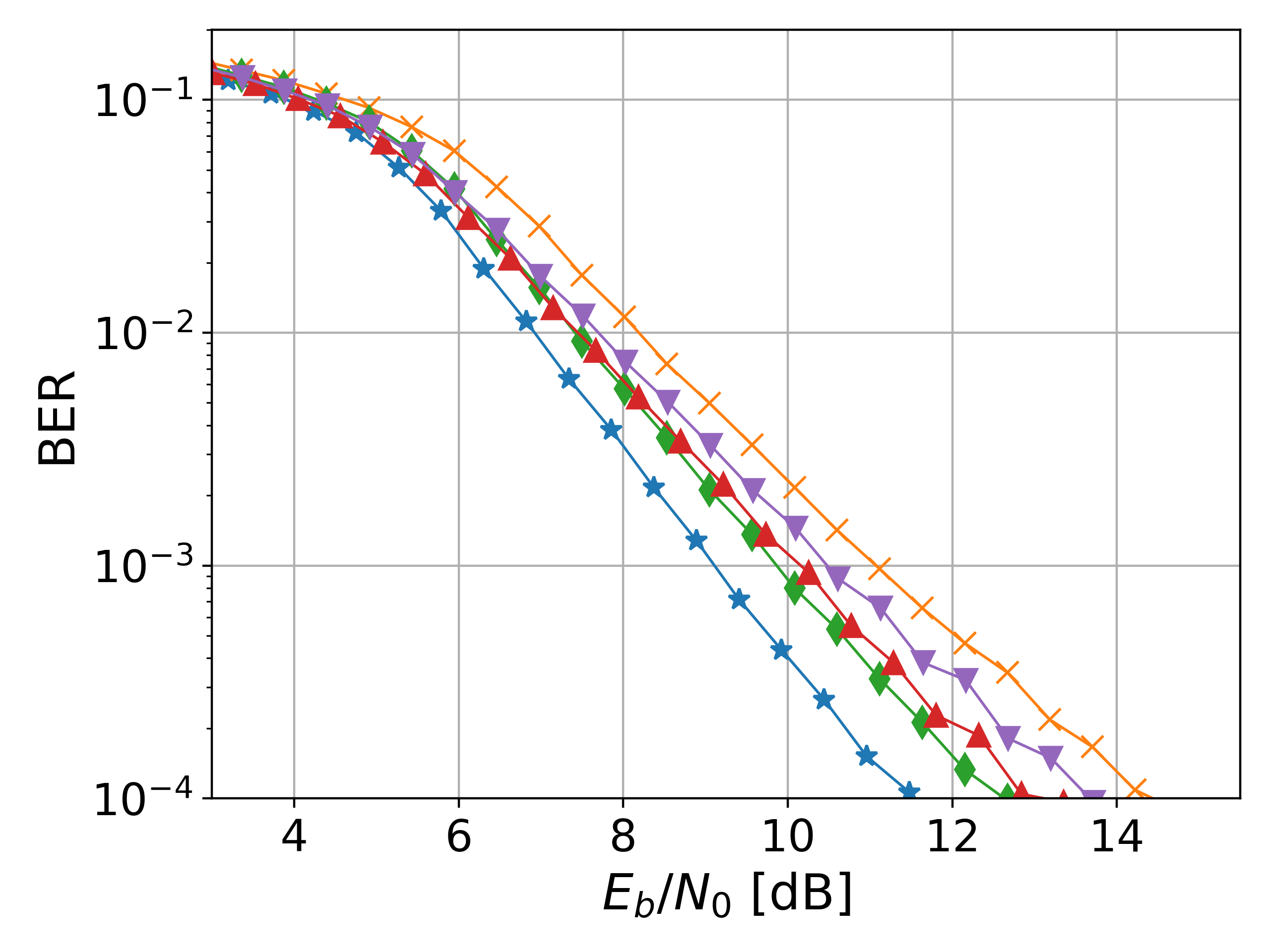}
         \caption{1P \& Low speed: 3.6\si{\km\per\hour}.\label{fig:ber_1p_low}}
     \end{subfigure}
     \begin{subfigure}[b]{0.3\linewidth}
         \centering
         \includegraphics[width=1.0\textwidth]{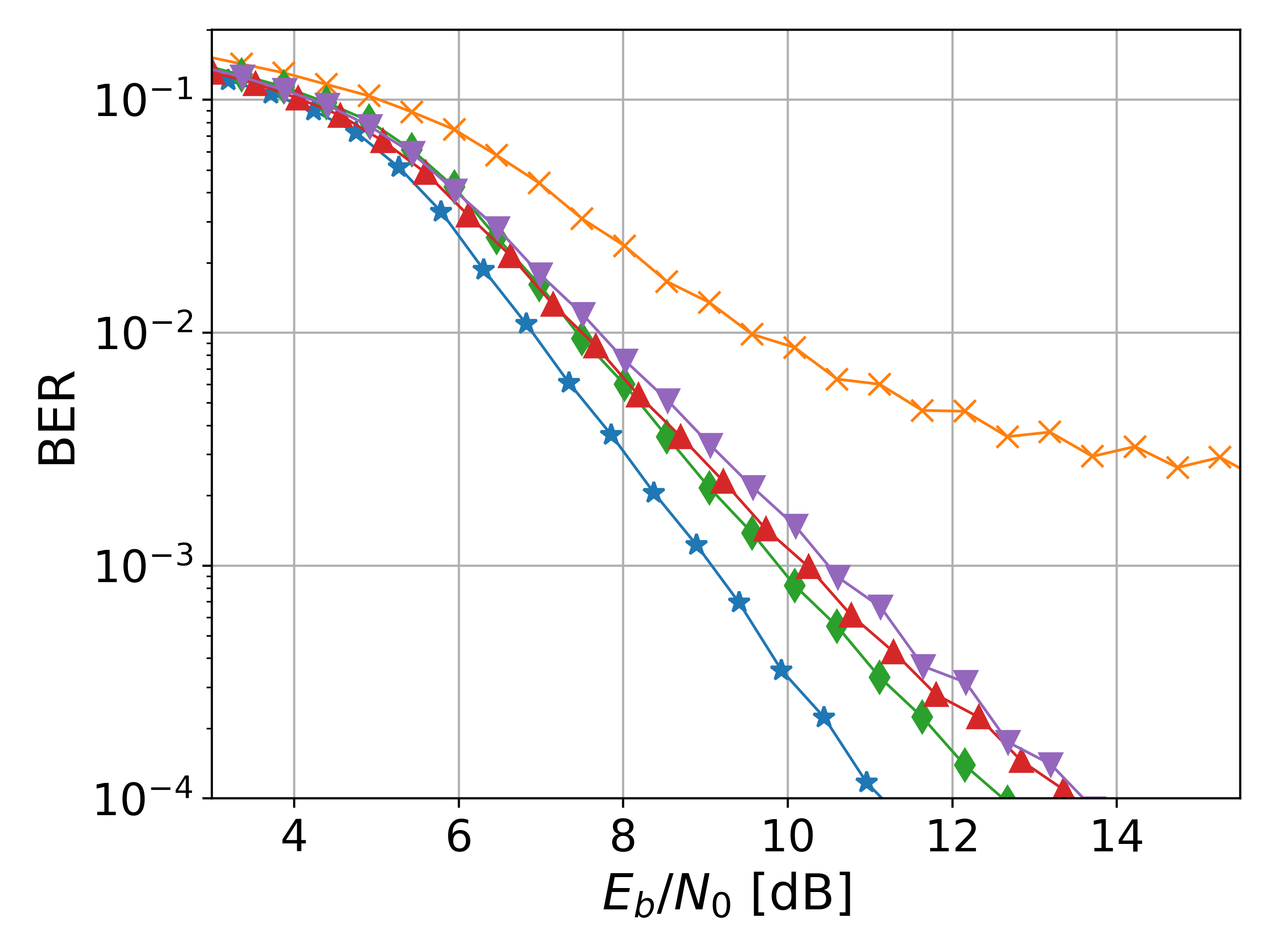}
         \caption{1P \& Medium speed: 36\si{\km\per\hour}.\label{fig:ber_1p_med}}
     \end{subfigure}
     \begin{subfigure}[b]{0.3\linewidth}
         \centering
         \includegraphics[width=1.0\textwidth]{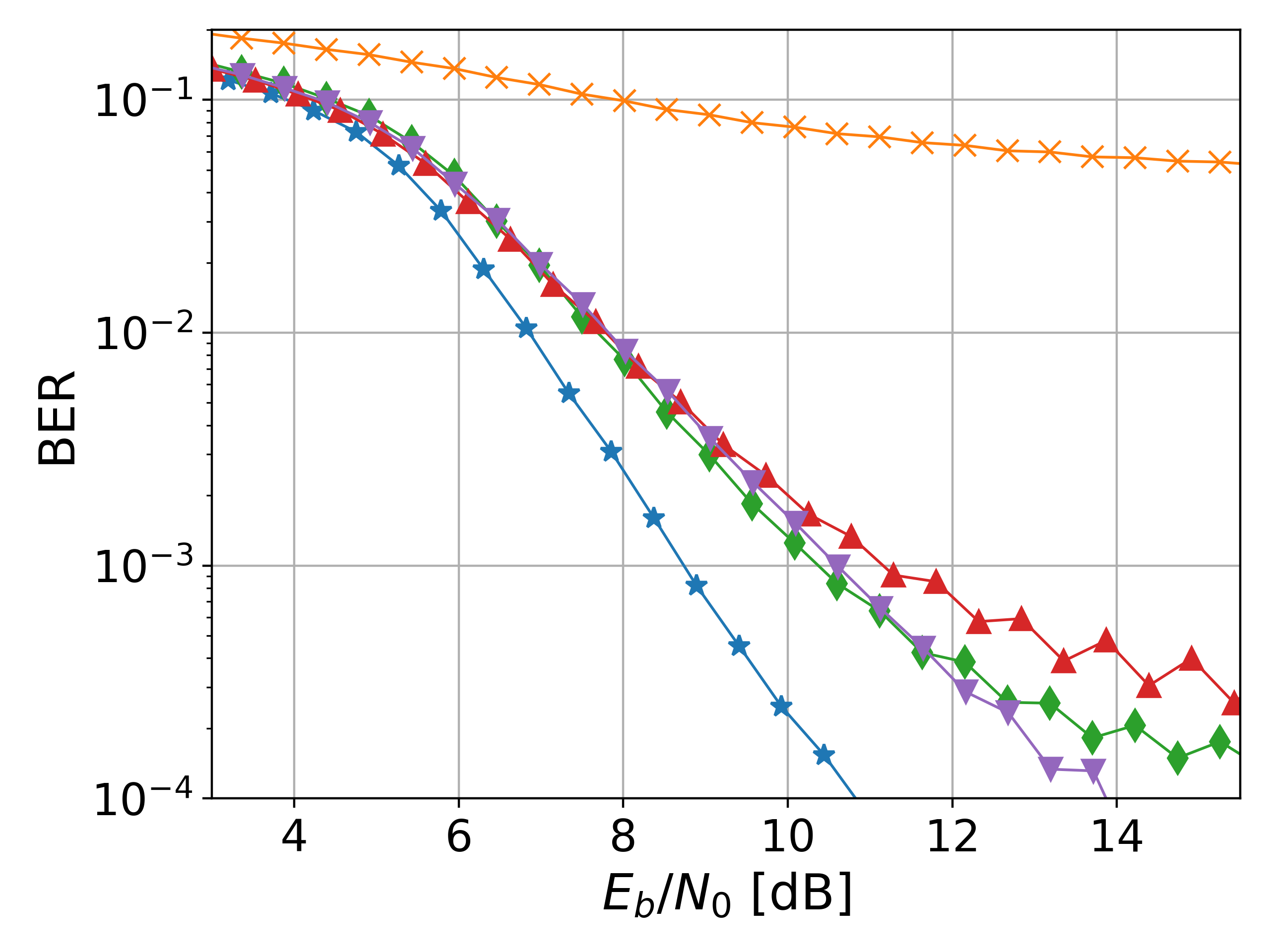}
         \caption{1P \& High speed: 108\si{\km\per\hour}.\label{fig:ber_1p_high}}
     \end{subfigure}
	
     \begin{subfigure}[b]{0.3\linewidth}
         \centering
		\includegraphics[width=1.0\textwidth]{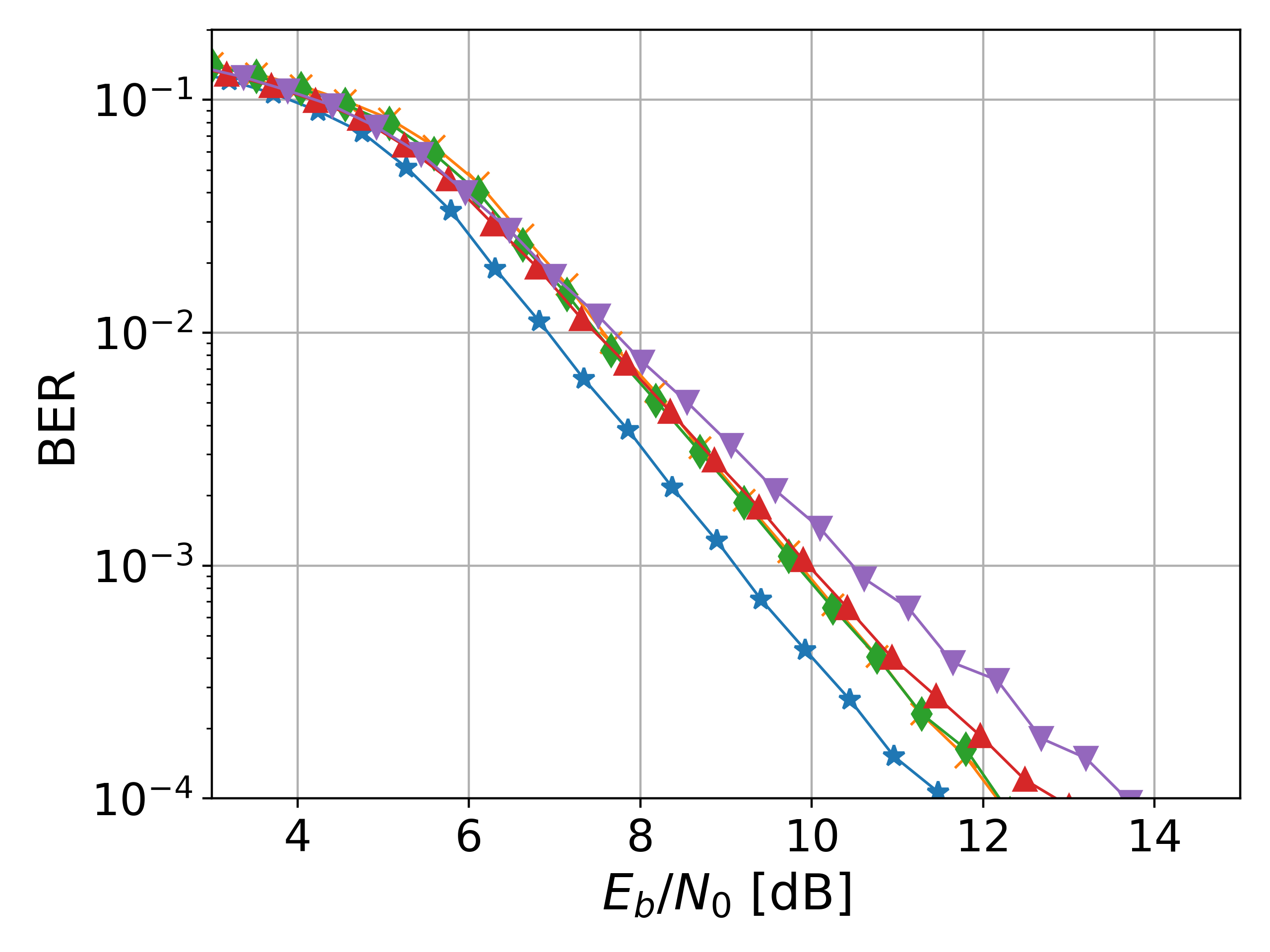}
         \caption{2P \& Low speed: 3.6\si{\km\per\hour}.\label{fig:ber_2p_low}}
     \end{subfigure}
     \begin{subfigure}[b]{0.3\linewidth}
         \centering
         \includegraphics[width=1.0\textwidth]{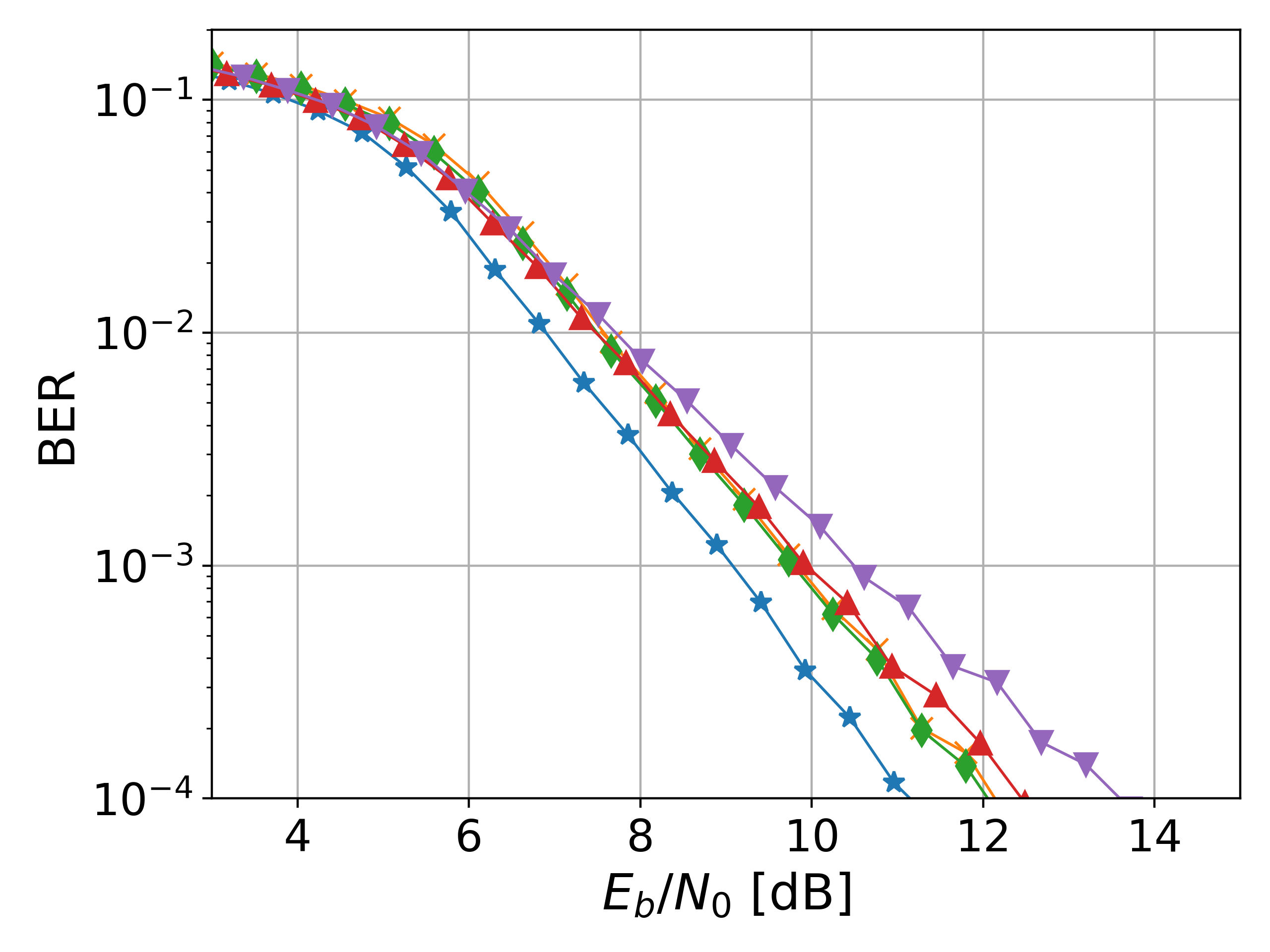}
         \caption{2P \& Medium speed: 36\si{\km\per\hour}.\label{fig:ber_2p_med}}
     \end{subfigure}
     \begin{subfigure}[b]{0.3\linewidth}
         \centering
         \includegraphics[width=1.0\textwidth]{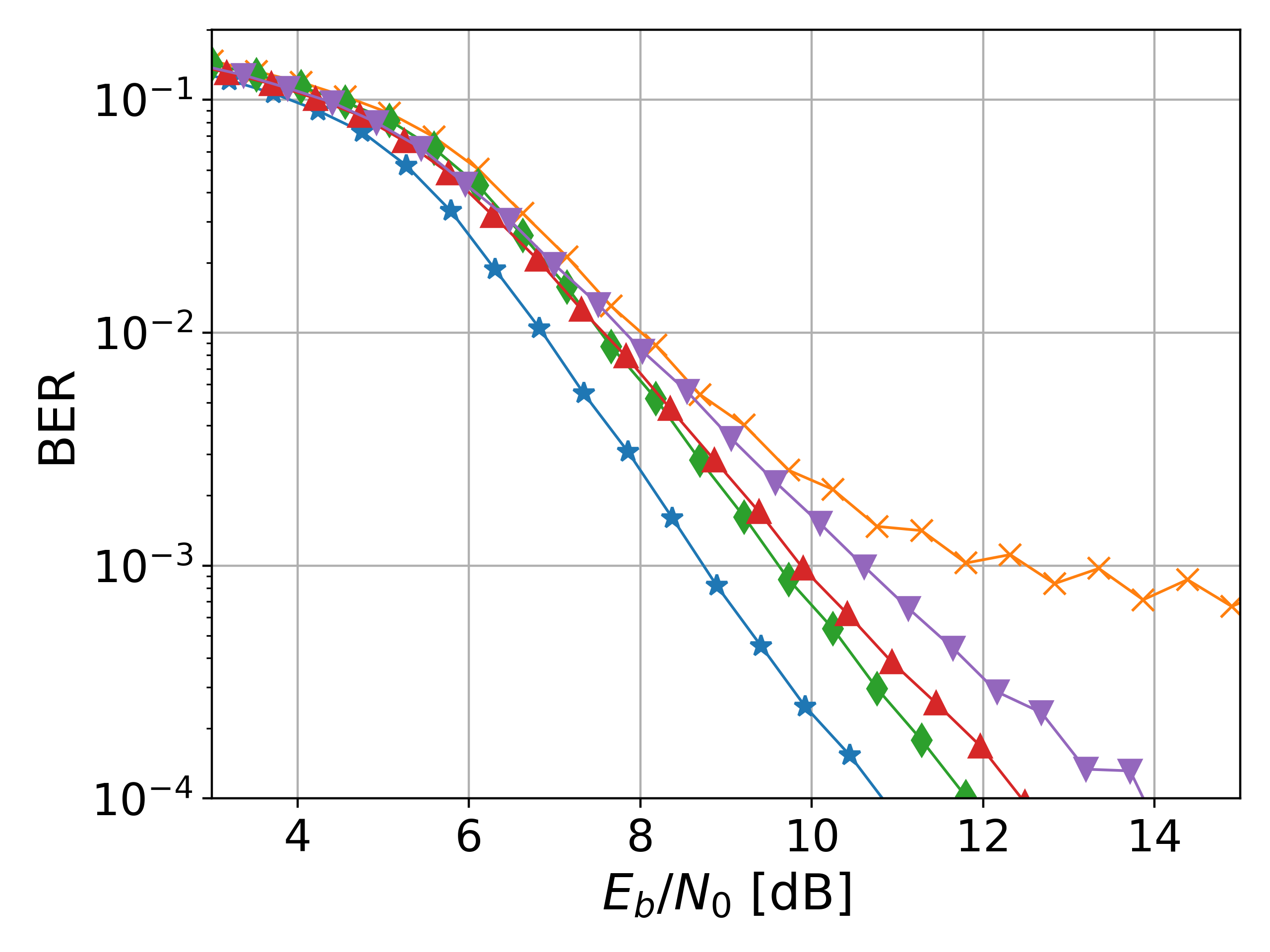}
         \caption{2P \& High speed: 108\si{\km\per\hour}.\label{fig:ber_2p_high}}
     \end{subfigure}
     \caption{BER achieved by the different schemes.\label{fig:ber}}
\end{figure*}

The channel model introduced in Section~\ref{sec:ch_mod} was considered to train the end-to-end system and to evaluate it against multiple baselines.
In addition to the \gls{LMMSE} baseline introduced in Section~\ref{sec:bsl}, a receiver that assumes~\eqref{eq:single_tap_transf} as transfer function and has perfect knowledge of $\gv$ was also considered.
A scheme that leverages the neural receiver from Section~\ref{sec:nn_arch} with \gls{QAM}, pilots, and \gls{CP} was also considered, as well as an approach that leverages the neural receiver and \gls{QAM} but without \gls{CP}.
Table~\ref{tab:setup_param} shows the parameters used to train and evaluate the machine learning-based approaches.
Training was done over the speed range from $0$ to \SI{130}{\km\per\hour}.
The \gls{3GPP} UMi LoS and NLoS channel models were considered, and the Quadriga~\cite{quadriga} simulator was used to generate a dataset used to train and evaluate the considered approaches.
The energy per bit to noise power spectral density ratio is 
\begin{equation}
\frac{E_b}{\sigma^2} \coloneqq \frac{1}{\rho}\frac{1}{mr\sigma^2}
\end{equation}
assuming $\EE\LP \abs{X_{i,k}}^2 \RP = 1$ for all \glspl{RE} $(i,k)$, and where $\rho$ is the ratio of symbols carrying data (the rest of the being used for pilots and \gls{CP}) and $r$ is the code rate.
Two pilot patterns from \gls{5GNR} were considered, which are shown in Fig.~\ref{fig:pp}.
The first one has pilots on only one \gls{OFDM} symbol (Fig.~\ref{fig:pp1}), whereas the second one has extra pilots on a second \gls{OFDM} symbol (Fig.~\ref{fig:pp2}), which makes it more suitable to high mobility scenarios.
These two pilot patterns are referred to as ``1P'' and ``2P'', respectively.

A standard \gls{5GNR} \gls{LDPC} code of length $1024\:$bit and with rate $\frac{2}{3}$ was used.
Decoding was done with conventional belief-propagation, using 40 iterations.
Each transmitted \gls{OFDM} frame contained 3 codewords and was filled up with randomly generated padding bits.
Interleaving was performed within individual frames.

The \gls{NN}-based receiver operates on the entire frame.
For fairness, the channel estimation~(\ref{eq:lmmse_est}) is also performed on the entire frame.
This involves the multiplication and inversion of matrices of dimension $1008\times1008$.
Moreover, it requires knowledge of the channel correlation matrix $\Rm$, which depends on the the Doppler and delay spread.
However, the receiver typically does not have access to this information, and assuming it to be available would lead to an unfair comparison with the \gls{NN}-based receiver which is not fed with the Doppler and delay spread.
Therefore, a unique correlation matrix has been estimated using the training set.
The so-obtained correlation matrix estimate was used to implement the \gls{LMMSE} baseline.

\subsection{Evaluation of end-to-end learning}
\label{sec:e2e_res}

\begin{figure*}
	\centering
	
     \begin{subfigure}[b]{\linewidth}
         \centering
		\includegraphics[width=0.9\textwidth]{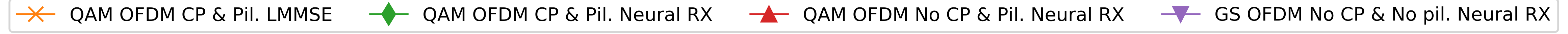}
     \end{subfigure}
	
     \begin{subfigure}[b]{0.3\linewidth}
         \centering
		\includegraphics[width=1.0\textwidth]{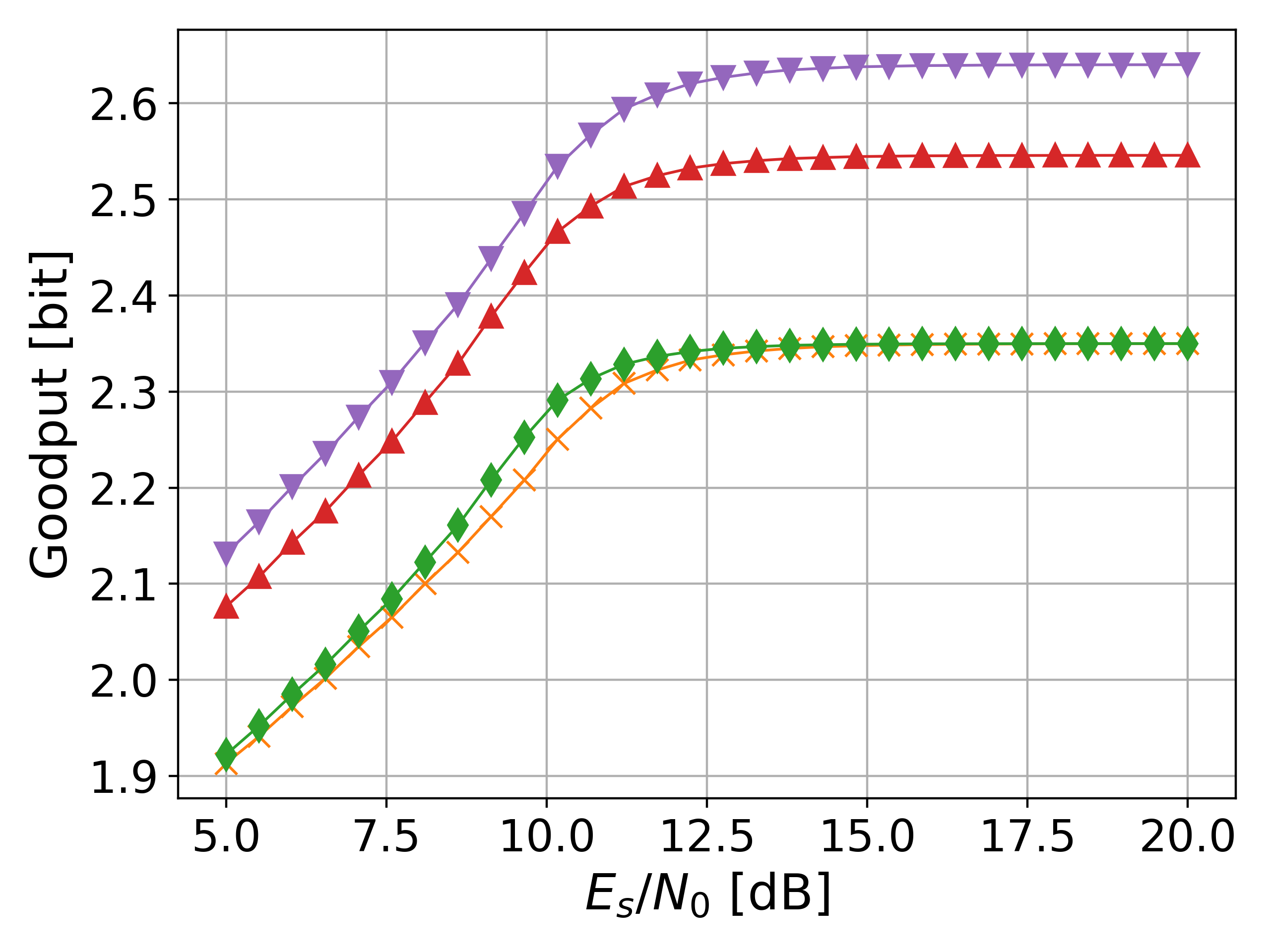}
         \caption{1P \& Low speed: 3.6\si{\km\per\hour}.\label{fig:gp_1p_low}}
     \end{subfigure}
     \begin{subfigure}[b]{0.3\linewidth}
         \centering
         \includegraphics[width=1.0\textwidth]{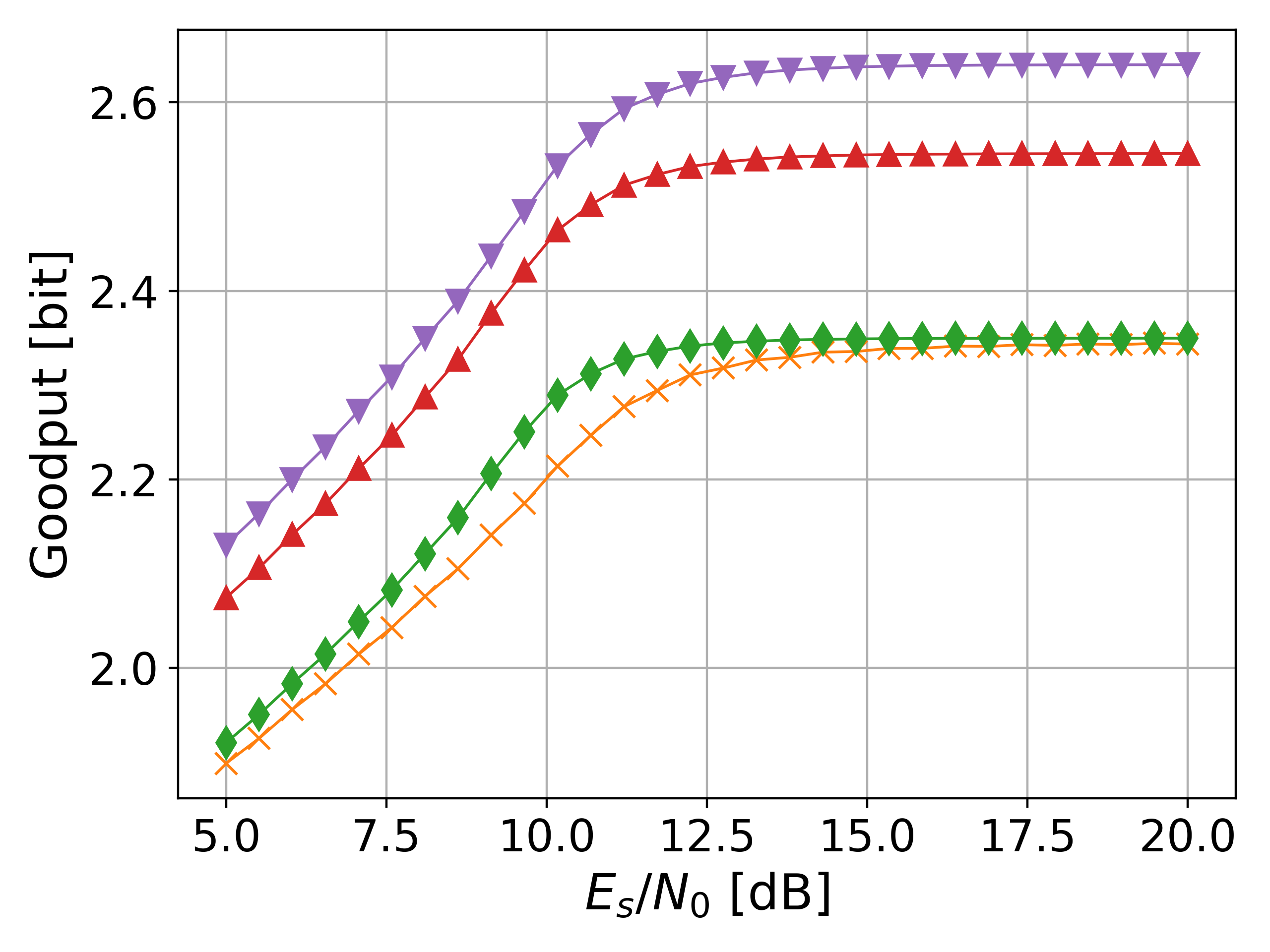}
         \caption{1P \& Medium speed: 36\si{\km\per\hour}.\label{fig:gp_1p_med}}
     \end{subfigure}
     \begin{subfigure}[b]{0.3\linewidth}
         \centering
         \includegraphics[width=1.0\textwidth]{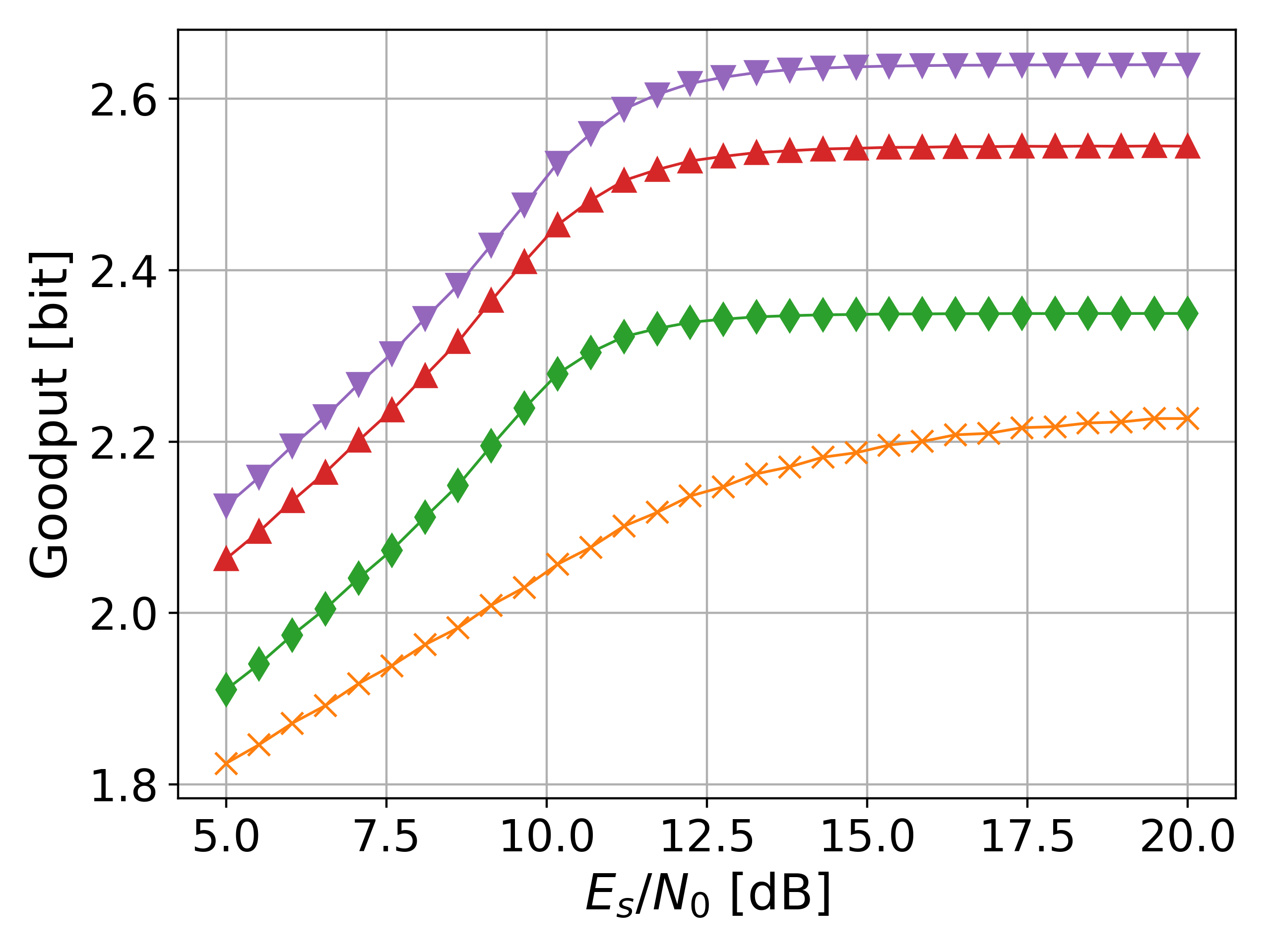}
         \caption{1P \& High speed: 108\si{\km\per\hour}.\label{fig:gp_1p_high}}
     \end{subfigure}
	
     \begin{subfigure}[b]{0.3\linewidth}
         \centering
		\includegraphics[width=1.0\textwidth]{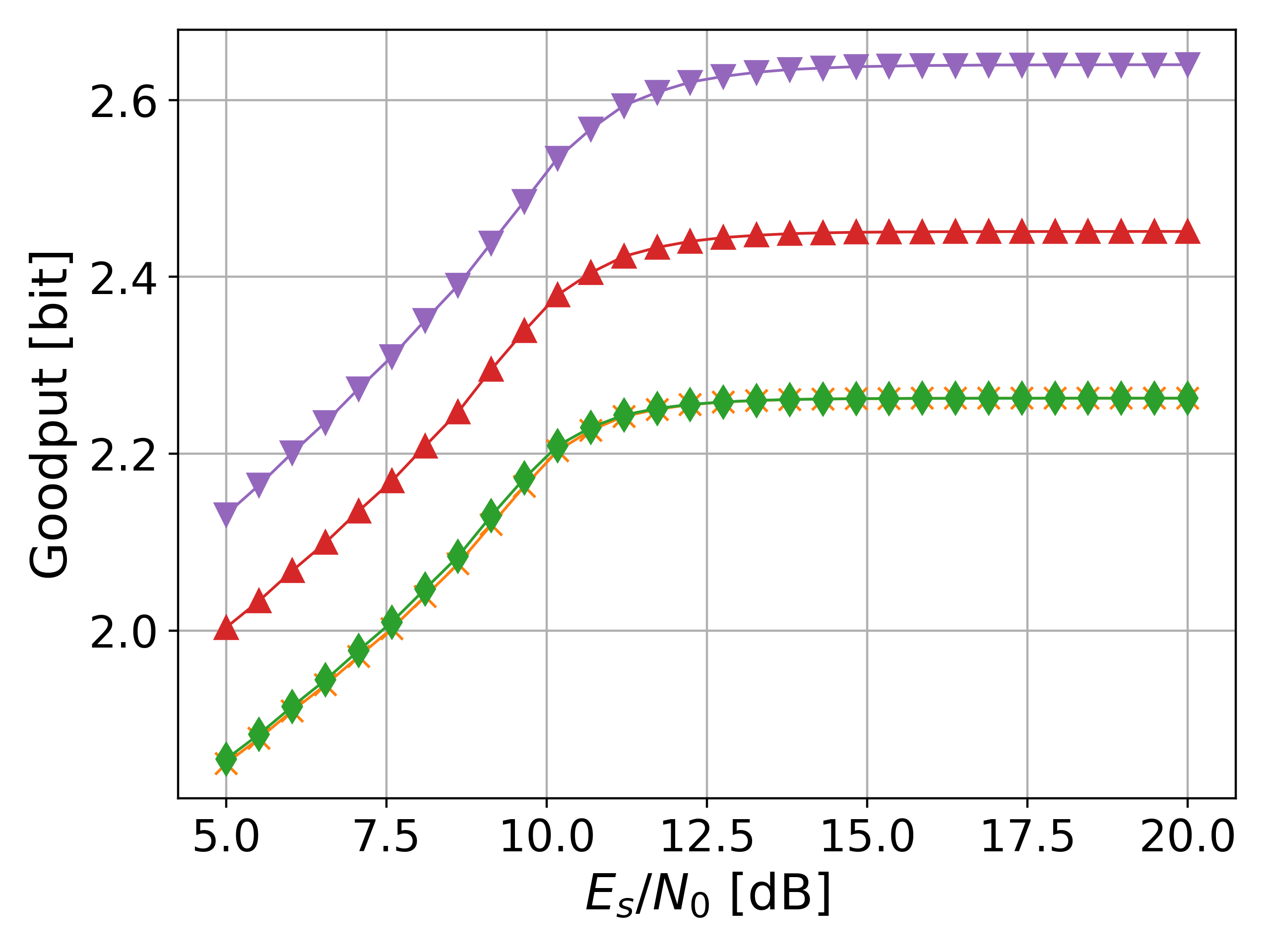}
         \caption{2P \& Low speed: 3.6\si{\km\per\hour}.\label{fig:gp_2p_low}}
     \end{subfigure}
     \begin{subfigure}[b]{0.3\linewidth}
         \centering
         \includegraphics[width=1.0\textwidth]{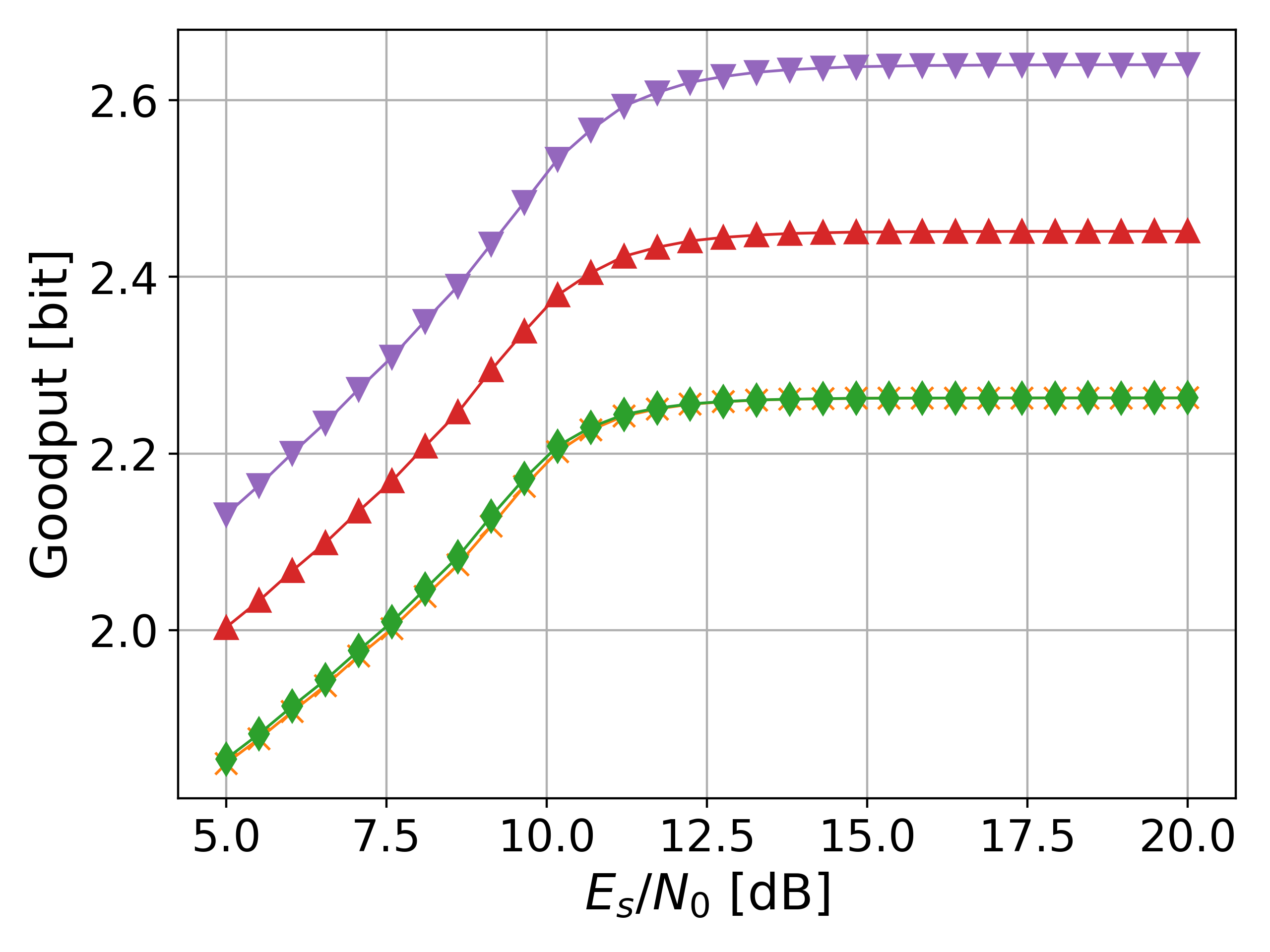}
         \caption{2P \& Medium speed: 36\si{\km\per\hour}.\label{fig:gp_2p_med}}
     \end{subfigure}
     \begin{subfigure}[b]{0.3\linewidth}
         \centering
         \includegraphics[width=1.0\textwidth]{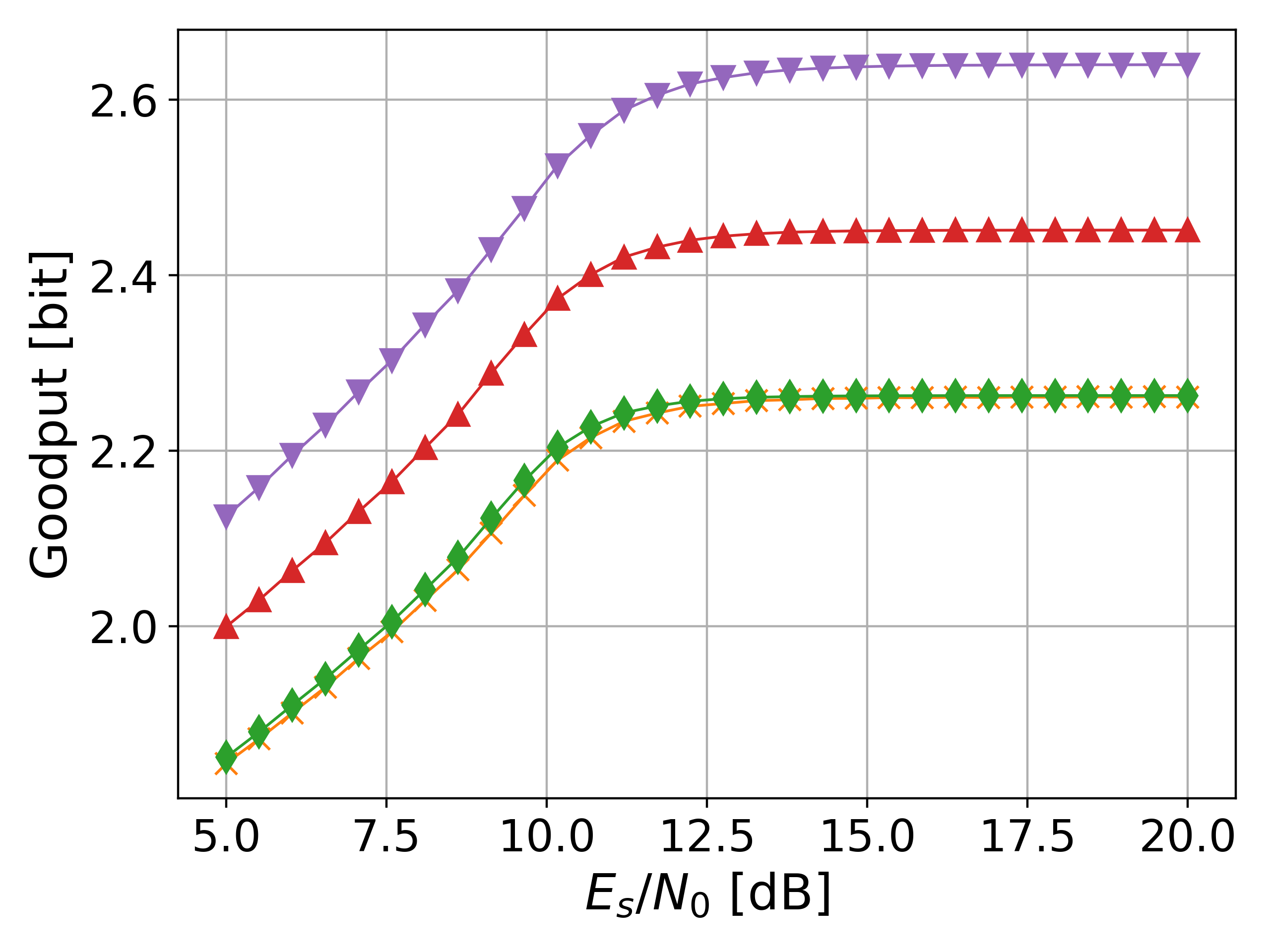}
         \caption{2P \& High speed: 108\si{\km\per\hour}.\label{fig:gp_2p_high}}
     \end{subfigure}
     \caption{Goodput achieved by the different schemes.\label{fig:gp}}
\end{figure*}

Fig.~\ref{fig:const} shows the constellation and labeling obtained by training the end-to-end system introduced in Section~\ref{sec:e2e}.
One can see that the learned constellation has a unique axis of symmetry.
Moreover, a form of Gray labeling was learned, such that points next to each other differ by one bit.

The first row in Fig.~\ref{fig:ber} shows the \glspl{BER} achieved by the various schemes when the 1P pilot pattern is used for three different velocities.
Note that the \gls{GS} scheme does not use any pilots.
One can see that the \gls{LMMSE}-based baseline leads to the highest \glspl{BER} for all speeds.
Moreover, its \glspl{BER} significantly worsen as the speed increases.
The lowest \glspl{BER} are achieved by the neural receiver with \gls{QAM}, pilots, and \gls{CP}.
The \gls{GS} scheme and the approach leveraging the neural receiver with \gls{QAM}, pilots, but without \gls{CP} achieve similar \glspl{BER} for the low and medium speeds, slightly higher than the one achieved by the neural receiver with \gls{QAM}, pilots, and \gls{CP}.
However, at high speed, the systems leveraging the neural receiver, \gls{QAM}, and pilots decline whereas the \gls{GS} scheme is more robust and achieves the lowest \glspl{BER}.

The second row in Fig.~\ref{fig:ber} shows the achieved \glspl{BER} when leveraging the 2P pilot pattern.
As expected, using more pilots enables the schemes using them to achieve lower \glspl{BER}, except for the \gls{LMMSE} baseline.
However, this is at the cost of extra pilots, leaving less resources to transmit data-carrying symbols.
The benefits of achieving low \glspl{BER} without the requirement of transmitting pilots and \gls{CP}, as allowed by the \gls{GS} scheme, is that it enables higher \emph{goodput}, as shown in Fig.~\ref{fig:gp}.
The baseline assuming perfect channel knowledge was omitted for clarity.
The goodput measures the average number of bits per symbols successfully received, and is defined by
\begin{equation}
\text{Goodput} \coloneqq r \rho m \LB 1 - \text{BER}\RB
\end{equation}
where $\rho$ is the ratio of data carrying symbols.
Moreover, because the goodput accounts for the unequal average number of information bits transmitted per symbol among the different schemes through the parameter $\rho$, it is plotted with respect to the energy per symbol to noise power spectral density ratio, defined as $\frac{E_s}{\sigma^2} \coloneqq \frac{1}{\sigma^2}$.
These plots show that the end-to-end approach enables at least \SI{18}{\percent} gains over the \gls{LMMSE}-based baseline (Fig.~\ref{fig:gp_2p_high}), and at least \SI{4}{\percent} gains over the neural receiver with \gls{QAM}, pilots, but no \gls{CP}.
One can see that for all the considered speeds and pilot patterns, the \gls{GS} scheme achieves the highest goodput.
The pilot and \gls{CP}-based schemes saturate at lower values as some of the symbols are not transmitting data.

\section{Conclusion}
\label{sec:conclu}
We have investigated the performance of end-to-end learning on top of \gls{OFDM}, and have shown that end-to-end learning enables multi-carrier communication without the need for \gls{CP} or pilots, and without significant loss of \gls{BER}.
This enables significant gains in throughput, as more resources are available for transmitting data.
Moreover, the learned system works for both \gls{LOS} and non-\gls{LOS}, and for a wide range of speeds and \glspl{SNR}.
Thus, apart from throughput gains, pilotless transmissions could remove the control signaling overhead related to the choice of the best suitable pilot pattern and \gls{CP}.

\bibliographystyle{IEEEtran}
\bibliography{IEEEabrv,bibliography}

\end{document}